\documentclass[conference]{IEEEtran}
\IEEEoverridecommandlockouts
\usepackage{cite}
\usepackage{amsmath,amssymb,amsfonts}
\usepackage[utf8]{inputenc}
\usepackage{algorithm}     
\usepackage{algpseudocode}
\usepackage{graphicx}
\usepackage{booktabs}
\usepackage{caption}
\usepackage{subcaption}
\usepackage{soul}
\usepackage{textcomp}
\usepackage{xcolor}
\usepackage{authblk}
\usepackage{amsthm}

\usepackage{algpseudocode}
\usepackage{algorithm}
\usepackage{float} 
\usepackage{amsmath} 
\newcommand{\cref}[1]{(\ref{#1})}

\def\BibTeX{{\rm B\kern-.05em{\sc i\kern-.025em b}\kern-.08em
   T\kern-.1667em\lower.7ex\hbox{E}\kern-.125emX}}

\DeclareSymbolFontAlphabet{\mathbb}{AMSb}
\begin{document}
\bstctlcite{IEEEexample:BSTcontrol}

\voffset=0.02in
\textheight=9.28in

\title{Just-in-time Restoration with Distributed Fiber Sensing in Metropolitan Optical Networks}
\author[1] {Sleman Mouammar}
\author[1]{\'Italo B. Brasileiro}
\author[2]{Andr\'e C. Drummond}
\affil[1]{Technische Universit\"at Braunschweig, Germany.}
\affil[2]{Department of Computer Science, University of Bras\'ilia, Brazil.}


\maketitle

\begin{abstract}
Distributed Fiber Sensing (DFS) leverages optical backscattering signals to predict failure events and enable just-in-time restoration in metropolitan optical networks, i.e., without optical amplifiers. In this paper, we study the effectiveness of proactive restoration based on DFS information in all-optical networks, while considering different sensing devices' capabilities. We evaluate whether restoration can be provisioned just-in-time before a failure happens, and its impact on key performance metrics, including the number of affected and suspended optical circuits, bandwidth blocking rate, and service downtime. Simulation results demonstrate that just-in-time restoration enabled by DFS with a prediction time capability of 15~ms can reduce circuit disruptions by more than 90\% compared to restoration without sensing and ensure optical service continuity in optical networks comparable to resource-intensive protection schemes, at a fraction of the spectral resources.
\end{abstract}


\section{Introduction} \label{sec:sdm-eon}

Distributed Fiber Sensing (DFS) is a known technique in optical networks, based on the fundamental sensing features of optical fiber links, including acoustic effects, vibration, temperature, and strain~\cite{wellbrock2023explore}. Distributed acoustic sensing (DAS), as a particular form of DFS, provides extraordinary awareness capabilities that allow the monitoring system to sense beyond the fiber cable, including objects and events surrounding it. The non-exhaustive list of events includes excavation, cable hauling in ducts, cable movement or perturbation, earthquakes, and cable strumming~\cite{lindsey:22}. As such, sensing devices support survivability by providing failure event prediction capabilities to network management before potential failures occur, within 5 to 7 minutes before the fiber cut \cite{mazur2024real}.

\begin{figure}[ht]
    \centering
    \includegraphics[width=1\linewidth]{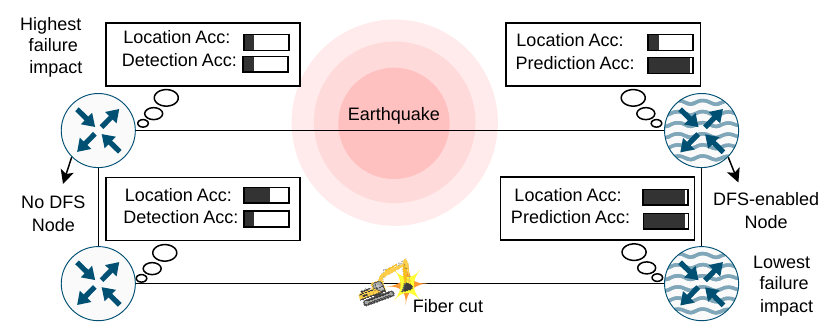}
    \caption{Metropolitan optical network with four optical links, two ``DFS-enabled'' and two ``no DFS'' nodes. Two failure events are demonstrated: a localized fiber cut with precise location, and an earthquake with wide coverage. Acc: accuracy. }
    \label{fig:sensingQuadrants}
\end{figure}

Traditional techniques for survivability in optical networks have been proven to reduce the impact of link failures, either by a simple switch to predefined backup or by a complete search and reallocation of new resources. While both protection and restoration solutions provide such alternatives, the restoration mechanism, in particular, may incur connection disruption following failures, and the post-failure reallocation process might be costly. Especially in high-capacity optical networks, where multiple users transmit in parallel at rates of hundreds of Gbps, failure prediction must occur as early as possible, and backup search must be fast and seamless. 

DFS devices deployed in optical nodes constantly measure channel state and signal quality information available in the backscattered light. Signal anomalies or unexpected variations occur right before failures, and the capture and processing of these occurrences provide insights into potential failures. Therefore, using sensing information extracted from the DFS devices, optical network protection 1:1 can timely switch circuits to backup paths before a failure occurs. Similarly, restoration solutions can operate proactively and search for new paths just-in-time (JIT) before failure occurrence ~\cite{765615}, thus reducing the impact of failure on service continuity. 

Fig. \ref{fig:sensingQuadrants} illustrates a metropolitan network with DFS. Nodes with sensing devices (i.e., DFS-enabled nodes) can predict in advance both localized and broader-scale failures. Whereas localized failures are more accurately predicted (bottom-right), larger disasters (an earthquake) cause lower localization accuracy (top-right). Nodes without DFS will suffer from delayed detection and usually rely on predefined solutions, protection 1+1, or reactive restoration. Therefore, applying survivability techniques combined with DFS mechanisms creates new challenges and new conceptual thinking about survivability. The open research question is hence whether service interruption can be eliminated for both techniques, and whether timely restoration can be performed on par with 1+1 protection (where both working and protection paths are active), i.e., without service interruption, while saving idle spare resources. 

Thus, a failure-aware and DFS-enabled network is capable of performing early event detection and localization with high accuracy, thus effectively predicting a failure. Under this assumption, and considering the occurrence of single failures in metropolitan networks, the contribution of this paper is threefold: i) a unified concept for node and control plane architectures of sensing-aware optical networks that integrates concepts as integrated sensing and communication (ISAC) and sensing control planes~\cite{szczerban2024optical}; ii) the evaluation through simulation of just-in-time restoration and protection under single link failures in sensed architectures, with key performance metrics such as the number of affected circuits, bandwidth blocking rate, and service downtime; and iii) the assessment of simulated sensing devices deployed with variable failure event prediction delays in metropolitan topologies, and the impact of JIT restoration based on the remaining time before failure.



Our results demonstrate that applying DFS across the network's nodes drastically  reduces circuit downtime and enhances service continuity. In particular, DFS devices with a prediction time of 15~ms before the failure deliver the most effective and efficient service maintenance.

The rest of the paper is organized as follows: Section \ref{sec:literature} presents the related work. Section \ref{sec:sensing} details our proposed architecture. Section \ref{sec:scenario} presents the evaluated scenario and the implemented techniques. Section \ref{sec:results} discusses the numerical results. Section \ref{Sec:Conclusion} concludes the paper.

\section{Related Work} \label{sec:literature}




Network survivability literature often reports Machine Learning (ML) techniques for failure event prediction. The work in \cite{10901306} proposes a Semantic FaultAware (SFA) architecture to enable proactive fault prediction for Wide Area Networks (WAN), by integrating a semantic layer above the control plane. Experiments based on data from a real network show that ML techniques allow the accurate prediction of potential network faults, achieving up to $88\%$ accuracy within 15 minutes of fault prediction time. Presuming that ML models are capable of successfully predicting failure events with high accuracy \cite{10901306}, the simulations in our work assume the deployment of such a trained model capable of failure prediction. This assumption enables us to evaluate sensing event detection on the network performance. However, any AI/ML-based technique can be used to further improve the results obtained here. 


Network protection might as well benefit from optical switching devices with high switching speeds. ITU-T standardizes protection mechanisms for service restoration within 50 ms, but Wavelength Selective Switches (WSS) are capable of providing reduced switching speeds of 10 to 20 ms. The impact of applying such devices for failure recovery is explored in \cite{he202550}, which proposes a shared protection heuristic for large-scale WDM networks with time-constrained resource allocation, aiming to recover paths within 50 ms. Our work explores the impact of devices with different reconfiguration times. We further consider the integration of DFS devices in protection scenarios, in order to verify that JIT enables proactive switching, achieving zero impact of failures.

The work in \cite{abdelli2022machine} shows that Optical Time-Domain Reflectometer (OTDR) traces often suffer from noise and signal misinterpretation, leading to false alarms. For more reliable monitoring, optical sensing systems should ideally incorporate a combination of Rayleigh, Raman, and Brillouin scattering to enhance accuracy and reduce false positives \cite{ezra:22}. Another work \cite{boitier:18} provides an EON testbed for JIT restoration that uses real-time State of Polarization (SOP) and ML as a fiber stress indicator. After detecting  stress data, the testbed starts rerouting before a fiber cut occurs, achieving protection-level switching times ($< 50$~ms) without allocating dedicated backup paths. In this paper, we investigate the  impact of applying various DFS devices with different failure prediction times (interval between a failure event prediction and its occurrence) on reallocation efficiency.

More recently, in \cite{szczerban2024optical}, a Fiber Sensing Control Device (FSCD) with a control plane that supports integration of sensing and communication is introduced. The work discusses the strategic placement of FSCDs to define sensing regions, combining both SOP and OTDR-based monitoring. The system presents local decision-making latencies below 50 ns and sensing responsiveness under $1~\mu $s, reinforcing the fast feedback sensing devices provide. In our paper, we assume that such sensing devices are deployed \cite{szczerban2024optical} and are distributed to provide total network coverage.

In \cite{shakouri2021proactive}, a JIT restoration solution is presented to counter single-link failures. The search for an alternative path happens just-in-time before failure, however, without resource reservation, and the solution is stored in a centralized control plane. The technique enhances QoS by replacing low-priority connections  with high-priority ones that are predicted to fail, reducing the allocation of free slots. This reduces the frequency of reallocation attempts, enabling faster and more efficient reallocation. In our paper, the search for the backup restoration path is conducted during the sensing-based detection phase. Upon finding free slots, circuits are reallocated before link failure, releasing the previously allocated resources. 


\section{Sensing-aware architecture} \label{sec:sensing}

\begin{figure*}[ht]
    \centering
    \includegraphics[width=0.9\linewidth]{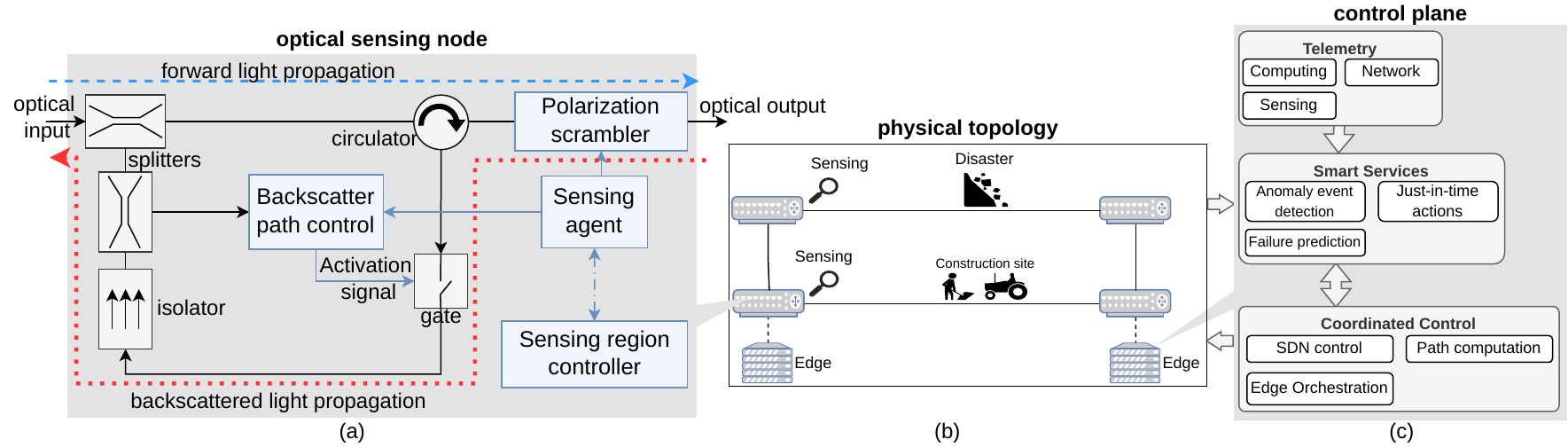}
    \caption{Sensing aware network with optical sensing node and control plane architectures.}
    \label{fig:sensingNode}
\end{figure*}

Turning the network into a global sensor allows management and control systems to leverage real-time environmental data from each path or link, enhancing the network's robustness and reliability. Fiber Sensing Control Devices (FSCD) are capable of sensing based on the SoP and backscattered light. Fig. \ref{fig:sensingNode} presents the architecture of a sensing-aware network~\cite{szczerban2024optical}.

In Fig. \ref{fig:sensingNode}(a), the optical signal enters the node via a coupler, which combines the incoming signal with the backscattered signal that comes from the output port. This facilitates transmitting backscattered light to the previous sensing nodes. The forward propagating light proceeds to a circulator, which directs it to the output port. A polarization scrambler affects both propagation directions, varying the SoP to avoid polarization-dependent effects. The backscattered light enters the node via the output port and is directed to a path controlled by the optical gate. A real-time sensing agent measures and controls both the SoP of the forward signal and the backscattered signal, translating and executing sensing policies defined by the sensing control plane.

Fig. \ref{fig:sensingNode}(c) illustrates components of the sensing control plane. The network data plane (Fig. \ref{fig:sensingNode}(b)) includes optical fibers, optical switching and routing infrastructure, as well as sensing infrastructure. We consider that a metropolitan network can be an elastic or classical WDM network able to manage the allocation and reallocation of circuits in response to failure events. There is a compute infrastructure collocated with the network infrastructure, implemented as edge computing sites of an optical network. In Fig. \ref{fig:sensingNode}(c), we envision that the corresponding optical control plane includes three main functional modules, i.e., Telemetry, Smart Services, and Coordinated Control functions. Telemetry interfaces with all optical and edge devices, including network, computing, and sensor equipment, to extract relevant data in a timely and granular manner, store it efficiently, and provide push/pull APIs for access. Both the fiber sensor and monitoring data can be gathered, including: (i) Link state: availability of the network links and packet-level information (rate, discards, errors); (ii) QoT metrics: signal to noise ratio (SNR), bit error rate (BER), Q-factor, error vector magnitude (EVM), and the shape of the eye diagram; (iii) Computing resources: edge servers memory/CPU/storage/application usage as well as their inter-connectivity (e.g., switches, ROADMS, passive interconnects); and (iv) Optical fiber sensor metrics.

The Smart Services functional module processes telemetry data, which, in addition to generating traditional network and computation performance metrics, can also leverage AI/ML sub-modules to extract meta-information. This enables the detection and prediction of anomaly events that could lead to faults, thereby identifying time-varying network vulnerabilities. This module includes key functions of the control plane and, most importantly, the Just-in-time actions to select the mitigation technique (e.g., reservation or restoration) to be applied in the network, based on the quality of the information provided by the detection and prediction sub-modules. 

Finally, the Coordinated Control module makes appropriate network and edge reconfiguration decisions in response to imminent failures, maximizing overall system reliability. It also performs network reconfigurations to ensure connectivity in the presence of failure. This architecture provides the foundation for a fault-aware closed control loop, enabling network management to exploit DFS and preemptively and just-in-time act to restore the service. 


\section{Just-in-time Restoration} \label{sec:scenario}

In this section, we explore the failure-aware network survivability, which considers accurate failure event prediction and the application of just-in-time techniques. More in detail, we describe how 1:1 dedicated path protection (DPP) and restoration (on-demand) can be implemented.
We assume that all network links are monitored with 100\% failure prediction accuracy; therefore, false alarms are not in the scope of this paper. Using unlimited sensing represents the best case, as our goal is to measure the impact of sensing on service disruption and recovery in single-failure scenarios. 


We consider a fault-aware control-plane-enabled metropolitan EON (Fig. \ref{fig:sensingNode}) that manages the allocation and reallocation of circuits in response to failures.
Two primary scenarios are evaluated: i) standard (no sensing), in which the reaction to failures happens only after physical disruption, and all circuits using the failed link suffer nearly immediate outages; and ii) just-in-time, in which DFS sensors embedded in the nodes detect anomaly events in advance, and preemptive actions can be taken. To enable simultaneous data and sensing transmission, a fixed frequency range is dedicated to the sensing paths \cite{szczerban2024optical}.

The impact of circuit reallocation signaling after a failure is quantified as the overall circuit downtime. As shown in  Eq. (\ref{eq: downtime}), the downtime consists of two main components:  the cumulative setup time (ST) for all allocation attempts (a), and the switching time (SWT) \cite{765615},
\begin{equation} 
   \footnotesize Downtime = a \cdot ST + SWT
   \label{eq: downtime}
\end{equation}
For DPP, the switching time $SWT_{DPP}$ is given by Eq. (\ref{eq:ProtSwitchingTime}):
\begin{equation}
\footnotesize \text{$SWT$}_{\text{DPP}}= F +n \cdot P + (n + 1) \cdot D + 2 \cdot m \cdot P + 2 \cdot (m + 1) \cdot D
\label{eq:ProtSwitchingTime}
\end{equation}
where $F$ is the failure event detection time, $n$ and $m$ are the number of hops in the primary and backup paths, respectively, $P$ is the propagation delay, and $D$ is the node processing delay. In restoration, on the other hand, the control plane searches for an alternative path after detecting the failure event. Every allocation attempt incurs a setup time, defined in Eq. (\ref{eq:setuptime}):
\begin{equation}
\footnotesize \text{$ST$}=  n \cdot P + (n + 1) \cdot D
\label{eq:setuptime}
\end{equation}
Due to the reactive nature of restoration, the switching time ($SWT_R$) only considers signaling in the broken path up to the failed link. Accordingly, this $SWT_R$ is given by Eq. (\ref{eq:pst_restoration}):

\begin{equation}
\footnotesize \text{$SWT_\text{R}$}= F + n \cdot P + (n + 1) \cdot D
\label{eq:pst_restoration}
\end{equation}



\begin{figure}[ht]
    \centering
    \includegraphics[width=0.74\linewidth]{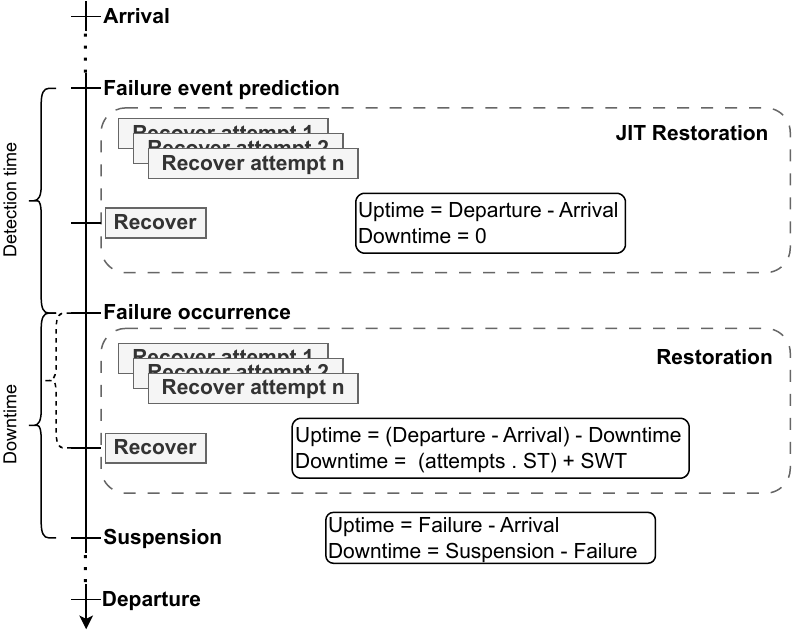}
    \caption{Circuit life-cycle for restoration.}
    \label{fig:circuitServiceTime}
\end{figure}

In optically sensed scenarios, once the DFS device detects an anomaly event, it notifies the control plane, enabling it to utilize the period between failure event prediction and occurrence to proactively search for alternative paths (ST) while the primary circuit remains active. Once the failure occurs, or after successfully establishing a new path, the control plane removes the primary circuit from the network to prevent transmission interruption and service unavailability (Algo.~\ref{alg:jit}, first case).
However, the switching penalty applies here if the path does not switch just-in-time before failure (Algo.~\ref{alg:jit}, second case). If a predicted failure event does not occur, the control plane will switch to the alternative path if successfully found; otherwise, the circuit remains in its primary path, so the impact remains negligible.  Fig. \ref{fig:circuitServiceTime} illustrates a circuit life-cycle with up- and downtime calculation. The main events are the circuit arrival, failure event prediction, failure occurrence, circuit suspension, and departure. We define Prediction time as the period between the failure event prediction and its occurrence. During this period, JIT restoration is attempted and, if successful, the circuit is reallocated (SWT) without service impact. After the failure, restoration is also attempted, within the suspension limit. 
The Downtime value can be: (i) zero, when no circuit disruption occurs or reallocation happens before failure occurrence; (ii) the time for its reallocation, in case of unsuccessful attempts before failure; or (iii) the suspension limit, if all reallocation attempts fail. The Just-in-time actions sub-module does this procedure in coordination with the Path computation and SDN control submodules (Fig.~\ref{fig:sensingNode}(c)).
\begin{algorithm}[ht]
\caption{JIT Restoration} \label{alg:jit}
\begin{algorithmic}[1]
\footnotesize
\algtext*{EndIf}
\algtext*{EndWhile}
\State AlternativePath $\gets$ NewPathSearch()
\State Attempt $\gets$ First
\While{FailureEventPrediction AND ! FailureOccurrence} \Comment{First case}
    \If{FirstAttempt  OR CircuitRemovalHappened}
        \If{AlternativePathSuccessfullyAllocated}
            \State SwitchPath(AlternativePath)
            \State RemovePrimaryCircuit()
        \EndIf
    \EndIf
\EndWhile

\State Attempt $\gets$ First
\While{FailureOccurrence AND ! Suspension} \Comment{Second case}
    \If{FirstAttempt OR CircuitRemovalHappened}
        \If{AlternativePathSuccessfullyAllocated}
            \State SwitchPath(AlternativePath)
        \EndIf
    \EndIf
\EndWhile
\end{algorithmic}
\end{algorithm}

\section{Numerical Results} \label{sec:performance}
\label{sec:results}

\begin{figure*}[ht]
    \centering
    \begin{subfigure}{0.40\textwidth}
    \centering
    \includegraphics[width=\linewidth]{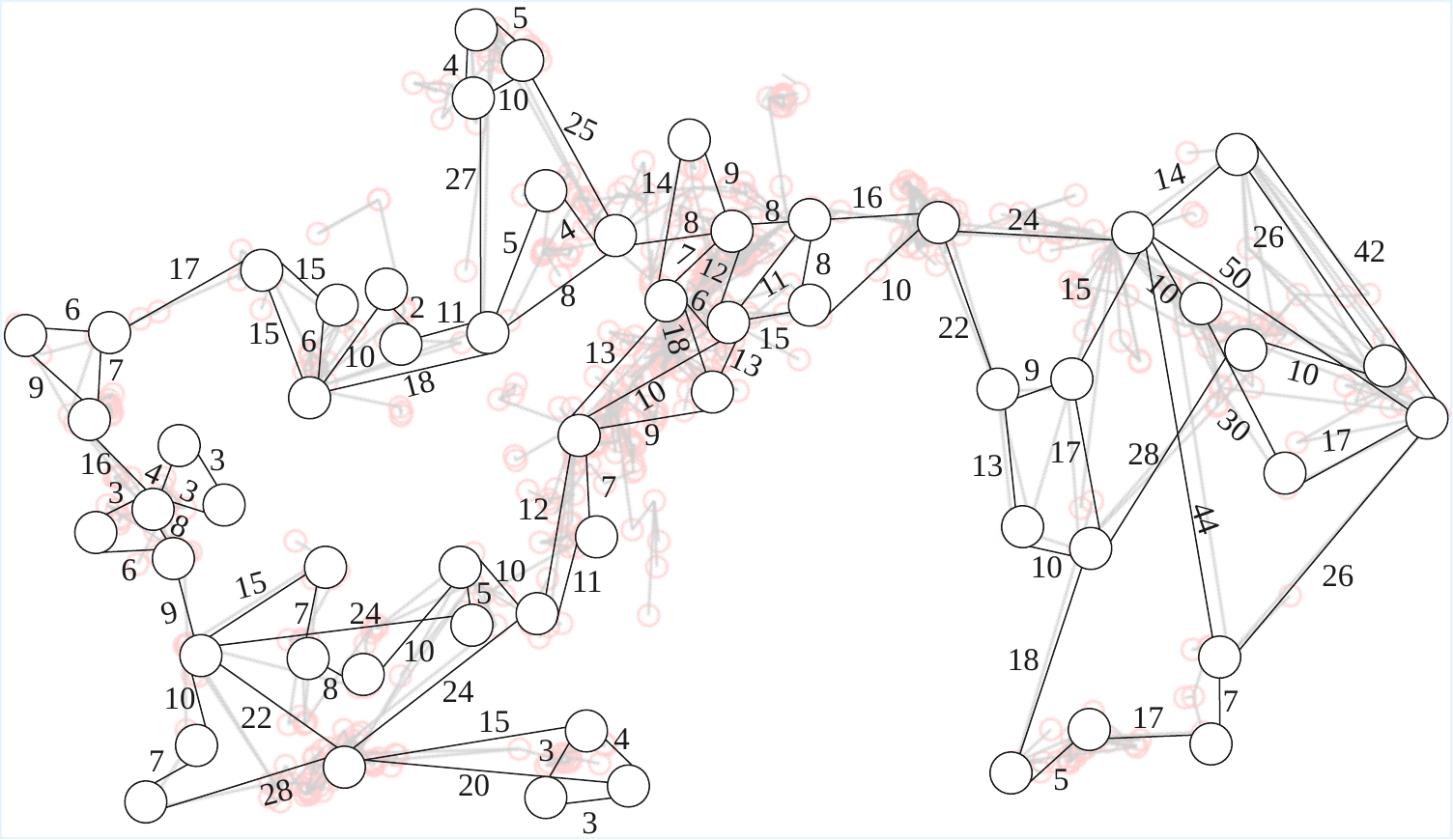}
    \caption{Catalunya topology \cite{catalunya}.}
    \label{fig: topology1}
    \end{subfigure}
    \begin{subfigure}{0.40\textwidth}
    \centering
    \includegraphics[width=0.83\linewidth]{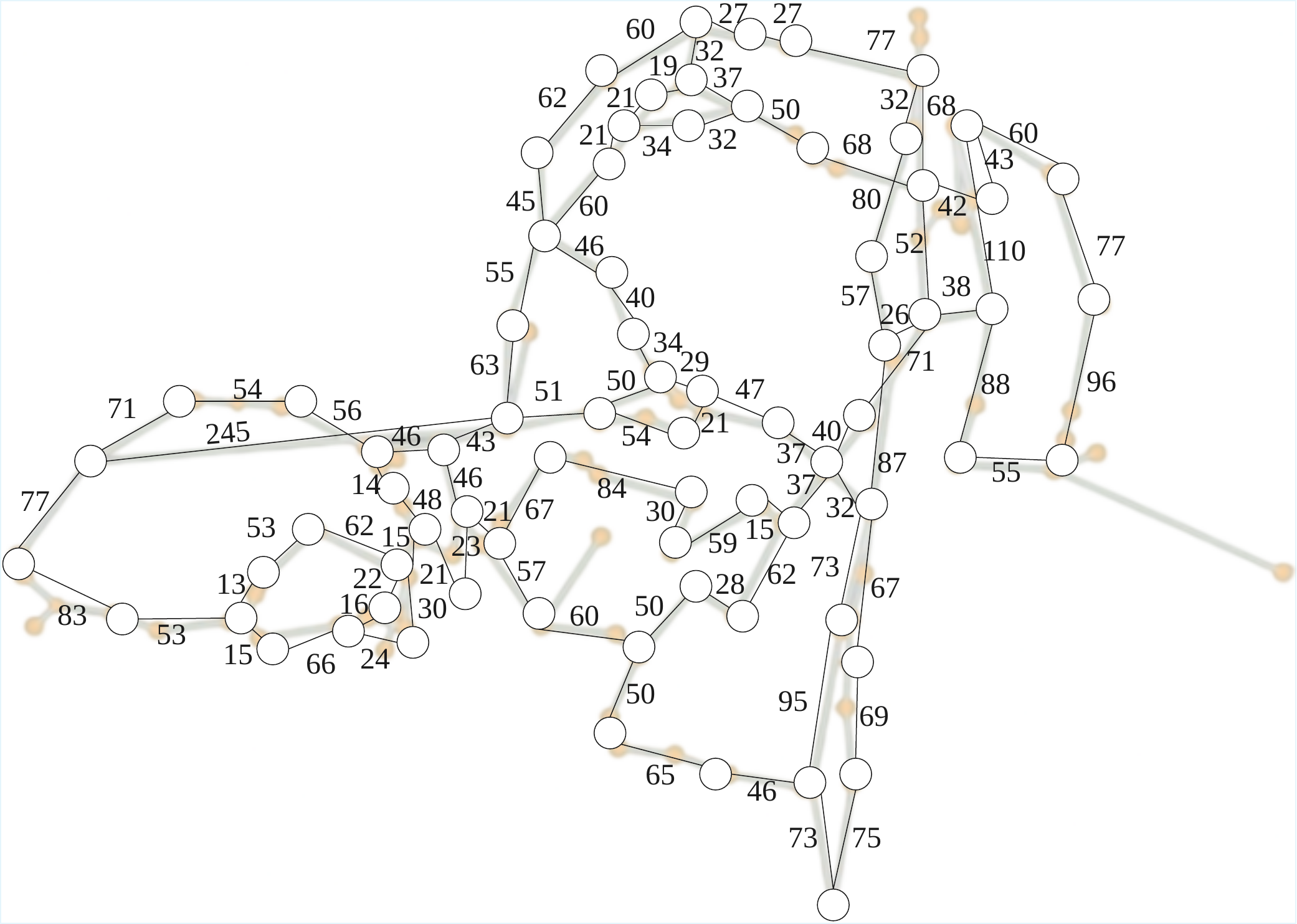}
    \caption{ION topology \cite{ion}.}
    \label{fig: topology2}
    \end{subfigure}

    \caption{The studied topologies with link lengths in km.}
\end{figure*}

We evaluate JIT restoration and compare it with traditional restoration and protection. We enforce that restoration requires searching for new paths and can have multiple attempts, which are executed before and after the failure (JIT restoration) or only after (restoration). Moreover, the implemented DPP guarantees circuit recovery and switches for the backup path either before failure (JIT protection) or after (protection).

We use an EON optical network simulator (https://ons-simulator.com/) to process the following events: service request, failure prediction, failure, and recovery. The results are shown with a 95\% confidence level. Each simulation runs 100,000 optical circuit requests, uniformly distributed among all node pairs, following a Poisson process. Each request has a mean holding time of 600s following a negative exponential distribution. Bandwidth requests vary uniformly from 1 to 400 Gbps, in steps of 1 Gbps. The available modulation formats are 8QAM and 16QAM with sub-carrier capacities of 37.5 and 50 Gbps, respectively. Their corresponding maximum transmission reaches are 1000 km and 500 km \cite{shakouri2021proactive}. The most efficient modulation is selected based on distance. Each link has 320 spectral slots, and a guard-band of one slot (12.5 GHz) is used between optical circuits. Route selection follows K-Shortest Paths ($k=3$) between all node pairs, and \textit{first-fit} policy is used for slot allocation.

The total simulation time was assumed to correspond to one year. Regarding failures, 100 uniformly distributed link failures were generated, with a mean time to repair (MTTR) of 4 hours. A 6-minute lead time was considered for failure event prediction \cite{mazur2024real}. We further evaluated how various DFS devices, each with distinct sensing capabilities, impact the results.
A maximum of 20 ms \cite{sambo2020locally} was considered for circuit reallocation attempts after a link failure.
The maximum link length for which sensing is applicable is 100 km, as no amplifiers are installed in the network (hence metro networks). The ST and SWT calculations follow Eqs. (\ref{eq:ProtSwitchingTime}), (\ref{eq:setuptime}), and (\ref{eq:pst_restoration}), considering D = 10~$\mu$s, P = 400~$\mu$s, and F = 500~$\mu$s \cite{765615} in scenarios without sensing. When sensing is enabled, an additional 340~ns response time from the sensing device is added to F \cite{szczerban2024optical}.

We simulate two topologies: the ION topology \cite{ion}, presenting a ring-composition metro network from New York City with 74 nodes with a $2.89$ average degree, and 92 bidirectional links, with an average length of $40.71$ km; and a simplified version of the Catalunya topology \cite{catalunya}, presenting a smaller and more connected ring-composition metro network from Catalunya City with 57 nodes, $2.96$ average node degree, and 86 bidirectional links of average $13.27$ km.



For the evaluation, we consider the following performance measures:  (i) Bandwidth blocking rate (BBR), the percentage of blocked circuits out of the total number of circuits requested; (ii) Affected circuits, the number of circuits that get disrupted by a failure event, regardless of whether they were successfully recovered after a failure or not; (iii) Suspended circuits, a subset of affected circuits that could not be recovered; (iv) Prediction time (ms), time between the failure event prediction and occurrence; and, (v) Downtime, overall unavailability of all circuits affected by a failure event.

\begin{figure*}[ht]
    \centering
    \begin{subfigure}{0.32\textwidth}
    \centering
        \includegraphics[width=\linewidth]{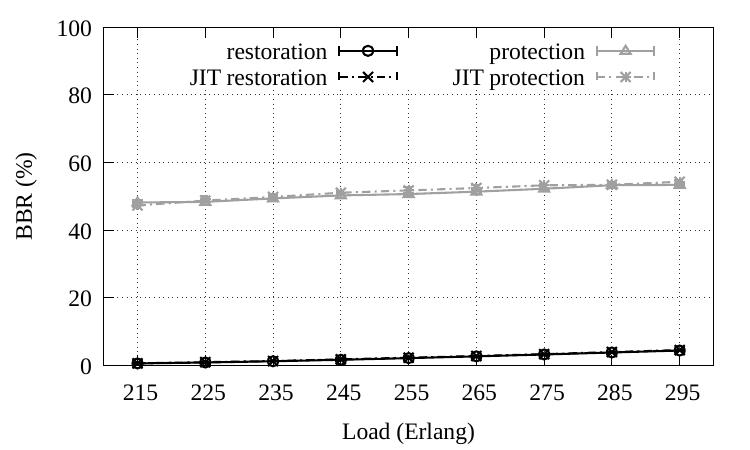} 
        \caption{BBR for different techniques.}
        \label{fig: BBR_ion}
    \end{subfigure}
    \begin{subfigure}{0.32\textwidth}
    \centering
        \includegraphics[width=\linewidth]{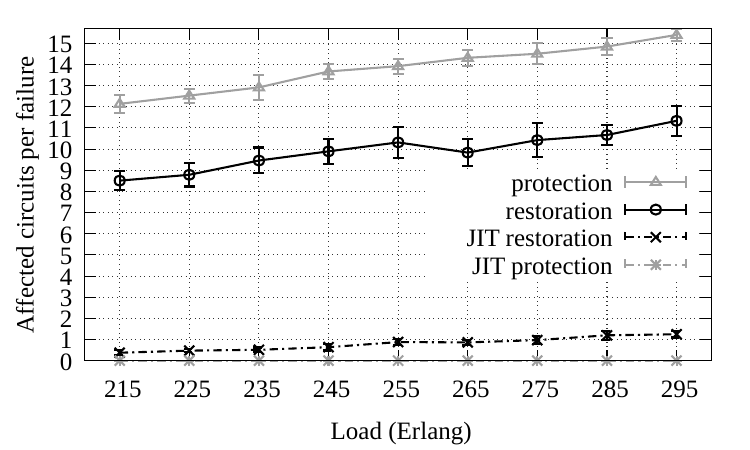}
        \caption{Average affected circuits per failure.}
        \label{fig:affectedCircuits}
    \end{subfigure}
    \begin{subfigure}{0.32\textwidth}
    \centering
        \includegraphics[width=\linewidth]{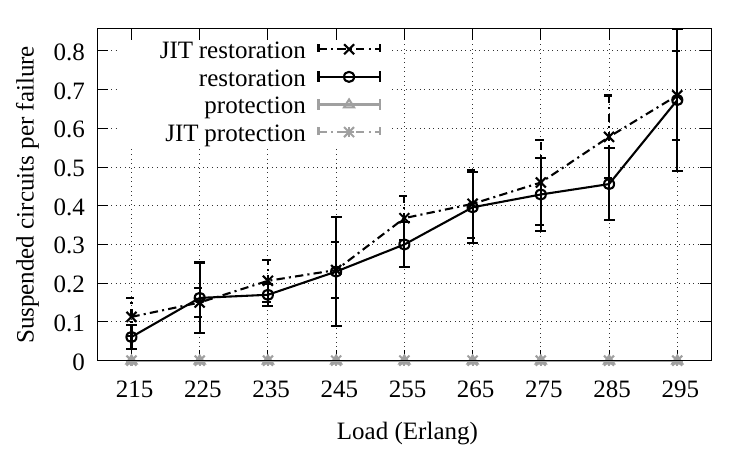}
        \caption{Average suspended circuits per failure.}
        \label{fig:suspendedCircuits}
    \end{subfigure}
    \caption{BBR, affected and suspended circuits performance with different survivability techniques.}
    \label{fig:Metricsx}
\end{figure*}

\begin{figure*}[ht]
    \centering
    \begin{subfigure}{0.30\textwidth}
    \centering
    \includegraphics[width=\linewidth]{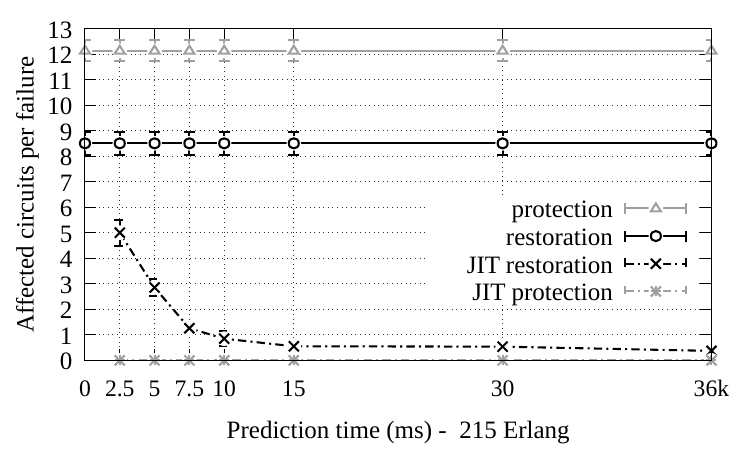}
    \caption{The impact on the affected circuits.}
    \label{fig:affectedBySensingTime}
    \end{subfigure}
    \begin{subfigure}{0.30\textwidth}
    \centering
        \includegraphics[width=\linewidth]{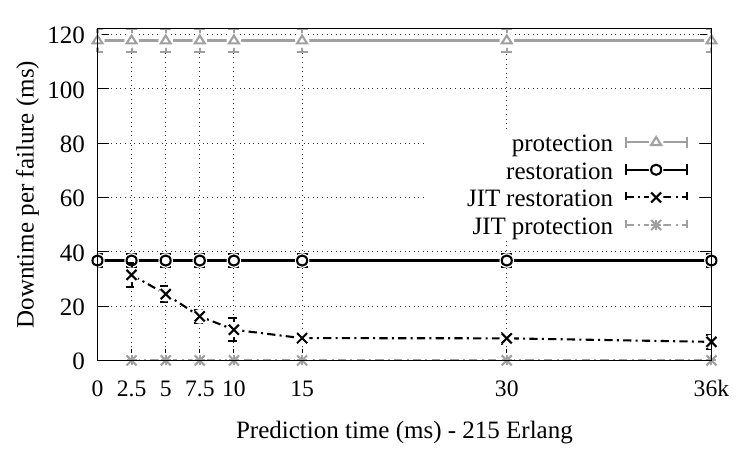}
        \caption{The impact on the downtime.}
        \label{fig:downTimePerFailure}
    \end{subfigure}
    \caption{Impact of deploying various DFS devices with different prediction times on the affected circuits and total downtime.} 
    \label{fig: DFSCapabilities}
\end{figure*}


We start by analyzing the BBR in Fig. \ref{fig:Metricsx}(a). These results exclude suspended circuits, which are evaluated later. The load values are chosen so that restoration BBR remains $<5\%$. As expected,  protection is highly inefficient ($\approx 50\%$ BBR) due to the dedicated allocation of backup resources, while restoration achieves blocking values $< 5\%$. Moreover, despite the fixed one-spectrum-slot dedicated to sensing in each link, the sensing overhead is negligible and does not affect the BBR.


We now evaluate the average number of affected circuits per failure event. As illustrated in Fig. \ref{fig:Metricsx}(b), protection results in more affected circuits due to its spectral overload, but ensures reallocation through the existing backup path. Moreover, JIT protection offers enough time to switch circuits from primary to backup path, neutralizing the impact of failures. Compared to restoration, JIT restoration technique significantly reduces the average number of circuit disruptions by $91.99\%$. However, network congestion still hinders the successful reallocation of a small subset of circuits. In particular, circuits traversing longer paths have a lower chance of successful reallocation, even when sufficient time is available for setup and switching before failure. For JIT restoration, an average of 0.79 circuits per failure cannot be reallocated, even with the extended time window provided by anticipated failure event prediction.
%


In Fig. \ref{fig:Metricsx}(c), we study the suspended circuits, which are those that could not be recovered within the suspension limit of 20 ms. Naturally, there are no such cases for protection, by definition. For the restoration, the results are equivalent for both scenarios. It is particularly interesting to note that for the JIT restoration, the average is $0.39$ suspended circuits per failure, which is roughly half of the number of affected circuits from Fig. \ref{fig:Metricsx}(c), which means that $50.6\%$ of the affected circuits are restored after the failure, thus accounting for downtime. 


Fig. \ref{fig: DFSCapabilities} shows the impact of DFS devices with different prediction times (defined in Fig. \ref{fig:circuitServiceTime}) on the number of affected circuits and total downtime.  Fig. \ref{fig: DFSCapabilities}(a) presents the average number of affected circuits as a function of failure event prediction time provided by different DFS devices. The upper curves represent the impact of protection and restoration without JIT allocation on the number of circuits affected by failure. Results indicate that a short prediction time ($\leq 15$~ms) significantly reduces the number of affected circuits. For instance, enabling 2.5~ms of prediction time for JIT restoration reduces circuit disruptions by $41.14\%$ compared to reactive restoration. As prediction time increases, the number of affected circuits declines, but stabilizes after roughly 15~ms. At this point, JIT restoration achieves a $93.5\%$ reduction (with $0.553$ affected circuits per failure) compared to reactive restoration ($8.507$). Beyond this point, additional prediction time yields negligible improvements (6 minutes suggested in \cite{mazur2024real}).

Fig. \ref{fig: DFSCapabilities}(b) evaluates the average total downtime per failure event. By increasing the prediction time, the overall downtime experienced by circuits with JIT restoration also reduces.
With JIT, a short sensing interval of 2.5~ms results in an average downtime reduction of $14.38\%$ (31.4 ms) compared to reactive restoration. The downtime decreases with longer prediction times, up to a reduction of $77.72\%$ (8.2 ms) at 15~ms. Beyond this point, the benefits from additional prediction time decline. The high downtime for 1:1 DPP is due to the total switching time across all affected circuits. This effect is nullified in JIT protection, as all affected circuits switch to their backups before failure occurs. For completeness, a prediction time of 6 minutes is considered to illustrate the impact of extended failure event prediction windows on circuit downtime, reaching an  $81.44\%$ reduction, and a downtime of 6.8 ms. 

Due to similar performance metrics obtained for both topologies, results for the Catalunya topology are presented only in Table \ref{tab:metrics}. These findings show that JIT restoration can achieve a level of efficacy considerably better than protection, and nearly equivalent to JIT protection, even predicting failure events 15 ms before they occur. Despite this limited prediction window, the available lead time is sufficient for the control plane to perform path recomputation and the associated signaling required for JIT restoration. Thus, the failure prediction capability provided by DFS devices enables preemptive restoration, leveraging real-time monitoring, while also reducing spectral overload caused by proactively allocated backup circuits in protection solutions.
\begin{table}
\centering
\footnotesize
\renewcommand{\arraystretch}{0.9} 
\begin{tabular}{l|cc|cc}
\textbf{Metric} & \multicolumn{2}{c|}{ION~~~~~Catalunya} & \multicolumn{2}{c}{ION~~~~~Catalunya} \\
\midrule
\multicolumn{1}{c}{} & \multicolumn{2}{c}{\textbf{JIT Protection}} & \multicolumn{2}{c}{\textbf{JIT Restoration}} \\
BBR (\%) & 51.33 &  54.48 & 2.35  & 3.17 \\
Affected circuits & 0.0 & 0.0& 0.79 & 0.97 \\
Suspended circuits& 0.0 & 0.0 & 0.35 & 0.46 \\
Downtime (ms) & 0.0 & 0.0 & 19.82 & 25.07 \\
\midrule
\multicolumn{1}{c}{} & \multicolumn{2}{c}{\textbf{Protection}} & \multicolumn{2}{c}{\textbf{Restoration}} \\
BBR (\%) & 50.79 &  53.46 & 2.26 & 3.04 \\
Affected circuits & 13.80 &  11.36 &9.91  & 8.13 \\
Suspended circuits& 0.0& 0.0 &0.32 & 0.39 \\
Downtime (ms) & 133.01  & 99.66 & 53.35 & 43.89 \\
\end{tabular}
\caption{Average metric values for different topologies.}
\label{tab:metrics}
\end{table}

\section{Conclusion and Future Work} \label{Sec:Conclusion}

This paper has shown how just-in-time (JIT) survivability methods can effectively reduce the circuits' total downtime and the number of affected circuits. Moreover, various sensing devices were simulated, and results showed that a 15~ms distributed fiber sensing (DFS) device window can significantly mitigate the impact of  network failures.  Furthermore, this study demonstrated that DFS-based restoration achieves effectiveness comparable to protection schemes while providing superior spectral efficiency. These findings open several avenues for future research, including investigations into multi-failure scenarios, the selection of edge nodes for control plane placement, optimized sensing device placement, and simulation of false alarms and missed detections.

\section{Acknowledgements}
The authors would also like to thank M.Sc. Iulisloi Zacarias,  Prof. Admela Jukan, and Prof. Thomas Schneider for their valuable contributions and revision of this work.

\bibliographystyle{IEEEtran}

\bibliography{mybib}
\end{document}